\documentclass[prl,10pt,twocolumn,superscriptaddress,aps,floatfix]{revtex4-1}
\usepackage{graphicx}
\usepackage{natbib}
\usepackage{url}
\usepackage{hyperref}
\usepackage[normalem]{ulem}
\usepackage{color}
\usepackage{mathtools,amssymb}
\usepackage{textcomp}
\usepackage{gensymb}
\usepackage[draft]{todonotes}   
\usepackage{braket}
\begin{document}

\title{Multiterminal Quantized Conductance in InSb Nanocrosses}
\author{Sabbir A. Khan*}
\affiliation{Microsoft Quantum Materials Lab Copenhagen, 2800 Lyngby, Denmark}
\affiliation{Center for Quantum Devices, Niels Bohr Institute, University of Copenhagen, 2100 Copenhagen, Denmark}

\author{Lukas Stampfer*}
\affiliation{Center for Quantum Devices, Niels Bohr Institute, University of Copenhagen, 2100 Copenhagen, Denmark}

\author{Timo Mutas}
\affiliation{Center for Quantum Devices, Niels Bohr Institute, University of Copenhagen, 2100 Copenhagen, Denmark}

\author{Jung-Hyun Kang}
\affiliation{Microsoft Quantum Materials Lab Copenhagen, 2800 Lyngby, Denmark}
\affiliation{Center for Quantum Devices, Niels Bohr Institute, University of Copenhagen, 2100 Copenhagen, Denmark}

\author{Peter Krogstrup}
\affiliation{Microsoft Quantum Materials Lab Copenhagen, 2800 Lyngby, Denmark}
\affiliation{Center for Quantum Devices, Niels Bohr Institute, University of Copenhagen, 2100 Copenhagen, Denmark}

\author{Thomas S. Jespersen}
\email{tsand@nbi.ku.dk}
\affiliation{Center for Quantum Devices, Niels Bohr Institute, University of Copenhagen, 2100 Copenhagen, Denmark}

\date{\today}

\begin{abstract}
By studying the time-dependent axial and radial growth of InSb nanowires, we map the conditions for the synthesis of single-crystalline InSb nanocrosses by molecular beam epitaxy. Low-temperature electrical measurements of InSb nanocross devices with local gate control on individual terminals exhibit quantized conductance and are used to probe the spatial distribution of the conducting channels. Tuning to a situation where the nanocross junction is connected by few-channel quantum point contacts in the connecting nanowire terminals, we show that transport through the junction is ballistic except close to pinch-off. Combined with a new concept for shadow-epitaxy of patterned superconductors on nanocrosses, the structures reported here show promise for the realization of non-trivial topological states in multi-terminal Josephson Junctions. 
\end{abstract}

\maketitle

Combining intrinsic confinement and high crystal quality, III-V semiconductor nanowires (NWs) have constituted an important experimental platform for mesoscopic physics and quantum devices for the past two decades \cite{Thelander:2006,Doh:2005,Nadj-Perge:2010,Hofstetter:2009}. Renewed interest was triggered by proposals for engineering exotic topological phases in strong spin-orbit interaction (SOI) one-dimensional (1D) NWs with proximity induced superconductivity \cite{Oreg:2010,Lutchyn:2010}. This led to significant theoretical works and experimental efforts in device engineering and material developments of hybrid semiconductor/superconductor structures \cite{Lutchyn:2018,Krogstrup:2015}. In addition, proposals for realizing quantum operations by braiding the world lines of non-abelian Majorana quasi-particles in networks of 1D hybrid nanowires \cite{Alicea:2011} create a need to extend the conventional linear NW platform towards branched hybrid structures. Different schemes are being developed towards planar NW networks \cite{Gooth:2017,Friedl:2018,krizek2018field, Vaitiekenas:2018, Aseev:2019} and vapor-liquid-solid NW growth has been extended to simpler branched structures either by changing the growth directions during growth \cite{Lao:2002, Dick:2004, plissard2013formation, krizek2017growth, Jespersen:2018} or by merging non-parallel NWs grown from tilted substrate facets \cite{dalacu2013droplet, Gazibegovic2017epitaxy}. The high mobility and strong SOI make indium antimony (InSb) the optimal candidate for transport measurements, however, branched InSb NW structures and nanocrosses have not so far been reported using molecular beam epitaxy (MBE) -- traditionally leading to crystals with the lowest impurity concentration.

Here, we demonstrate controlled synthesis of InSb nanocrosses (NCs) using MBE, where the challenge of initiating InSb growth is overcome using a two-step procedure, beginning with a short segment of InAs supporting the subsequent InSb NW. A NC geometry is enabled by growing NWs from facing, non-parallel $\langle 111 \rangle$ facets. We map the conditions for NC formation in terms of relative catalyst position, axial growth rates, and radial growth of InSb NW. Structural characterization using transmission electron microscopy (TEM) confirms a single-crystal Zinc-Blende (ZB) phase across the InSb NC. Further, we study the low-temperature electron transport of the InSb NCs with local gate-control of each individual terminal. Each terminal exhibits quantized conductance, and from analysis of the combined transport through two quantized constrictions we infer ballistic transport also over the NC junction, except for global gate potentials close to pinch-off. Finally, we demonstrate a new concept for merging the NC geometry with pre-defined substrate structures to enable \textit{in situ} patterning of epitaxial superconductors, which have been shown to substantially enhance the performance of hybrid devices \cite{khan2020highly, Carrad:2020, Heedt:2020}.

\begin{figure}[t!]
\vspace{0.2cm}
\includegraphics[scale=0.92]{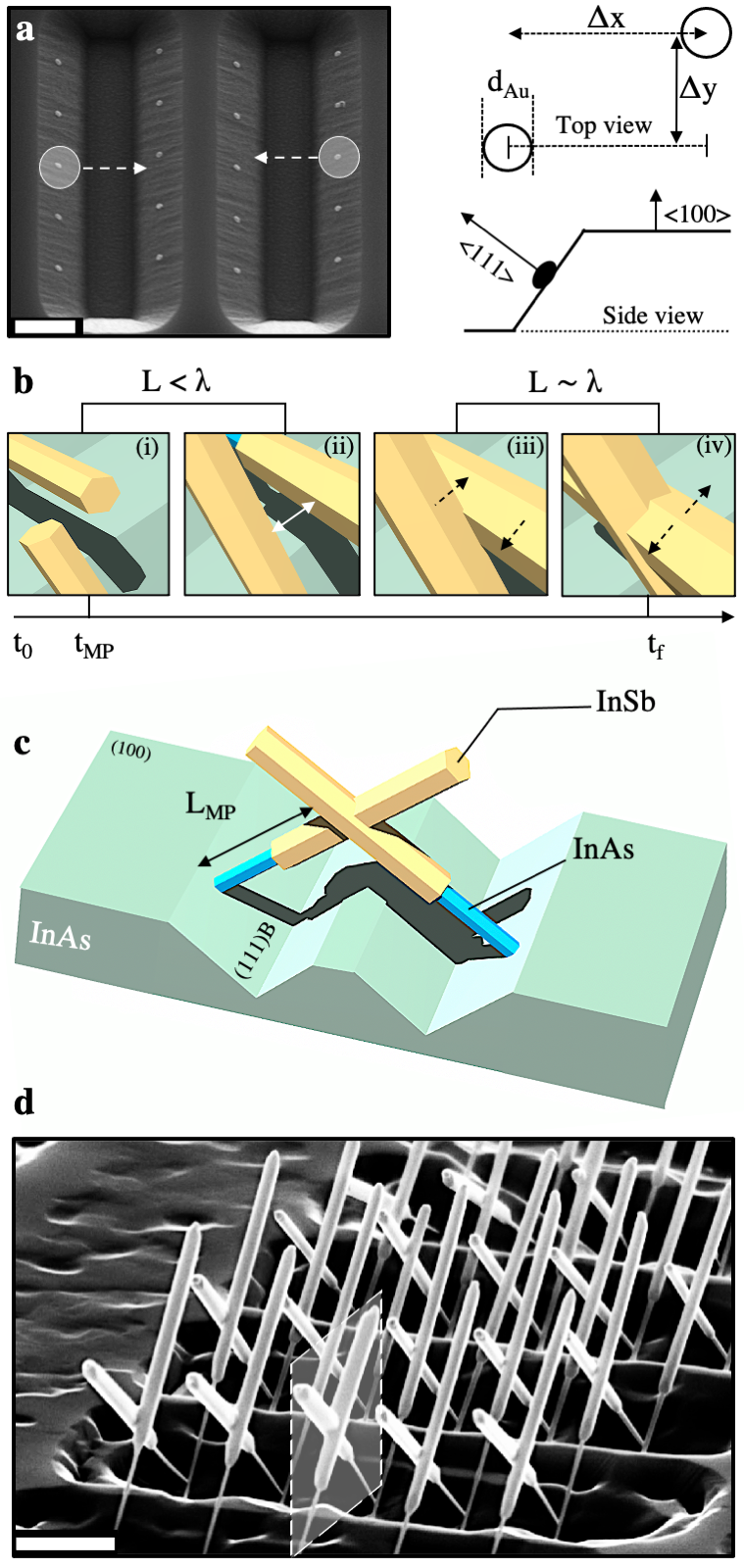}
\vspace{0.2cm}
\caption{\textbf{Substrate design and InSb nanocross formation.} \textbf{a}, Tilted scanning electron micrograph (SEM) of InAs (111)B trenches with deposited Au catalyst particles. The schematics illustrate the notation used for describing the layout. \textbf{b}, Schematics of the evolution of nanowires to merge and create nanocrosses. In the time scale, growth starts at $\mathrm{t_{0}}$, nanowires pass each other at $\mathrm{t_{MP}}$ and $\mathrm{t_f}$ is the final growth time. When $\mathrm{L< \lambda}$ (step (i) and (ii)), the radial growth is limited and nanowires pass each other without merging. When $\mathrm{L} \sim \lambda$ (step (iii) and (iv)), the radial growth overcomes $\Delta$y spacing and nanowires merge from the side-facets. Hence, radial growth from $\mathrm{t_{MP}}$ to $\mathrm{t_f}$ determines the nanocross formation. \textbf{c}, Schematic of the final nanocross structure. \textbf{d}, Growth demonstration of InSb nanocrosses from the identical substrate design shown in panel (\textbf{a}). Scale bars in (\textbf{a}) and (\textbf{d}) are is 1 $\mu$m.}
\label{fig1}
\end{figure}


\section{Results and Discussion}

\textbf{Growth of InSb Nanocrosses.} The growth of InSb NCs was enabled using InAs (100) substrates containing (111)B faceted trenches, which were fabricated by chemical etching following along the lines of Ref.\ \cite{khan2020highly}. Au catalyst particles with relative lateral distances $\Delta x$, $\Delta y$ were defined on opposing facets using electron beam lithography. See Fig.\ \ref{fig1}\textbf{a} for a definition of the coordinate system and a scanning electron microscope (SEM) micrograph of a typical substrate before growth. The result of the final NW structure depends critically on the relation between $\Delta$y, the diameter $d_\mathrm{Au}$ of the supersaturated Au catalyst that defines the initial diameter $d(t_0)$ of the InSb NW during axial growth, and the amount of radial growth when the NW reach to it's final diameter $d(t_\mathrm{f})$. For $\Delta \mathrm y < d_\mathrm{Au}$ the NWs meet “head-to-head” and the axial growth is interrupted preventing the formation of four-terminal crosses \cite{de2019crossed, kang2017wurtzite}. This situation is discussed in the Supporting Information (S1-S6). For $\Delta \mathrm y$ exceeding the final NW diameter $d(t_\mathrm f)$, the two NWs remain separated, while for a finite off-set $d_\mathrm{Au} <\Delta \mathrm y < d(t_\mathrm f)$ the resulting structure and the possibility of forming connected four-terminal crosses depends on the relation between $\Delta y$, $d_\mathrm{Au}$, $d(t_\mathrm{MP})$, and $d(t)$. Here, $t_\mathrm{MP}$ refers to the time when the two catalyst particles pass each other at the closest point. The sequence is illustrated in Fig.\ \ref{fig1}\textbf{b}. Ideally, the growth front (the size of which is given by $d_\mathrm{Au} = d(t=0)$) of the two NWs pass each-other with minimal separation thus, continuing axial growth while the simultaneous radial overgrowth eventually merges the two NWs epitaxially at the meeting-point. The final structure is schematically shown in Fig.\ \ref{fig1}\textbf{c} and a typical example of MBE grown NCs is presented in Fig.\ \ref{fig1}\textbf{d}.

\begin{figure*}[ht!]
\vspace{0.2cm}
\includegraphics[scale=1]{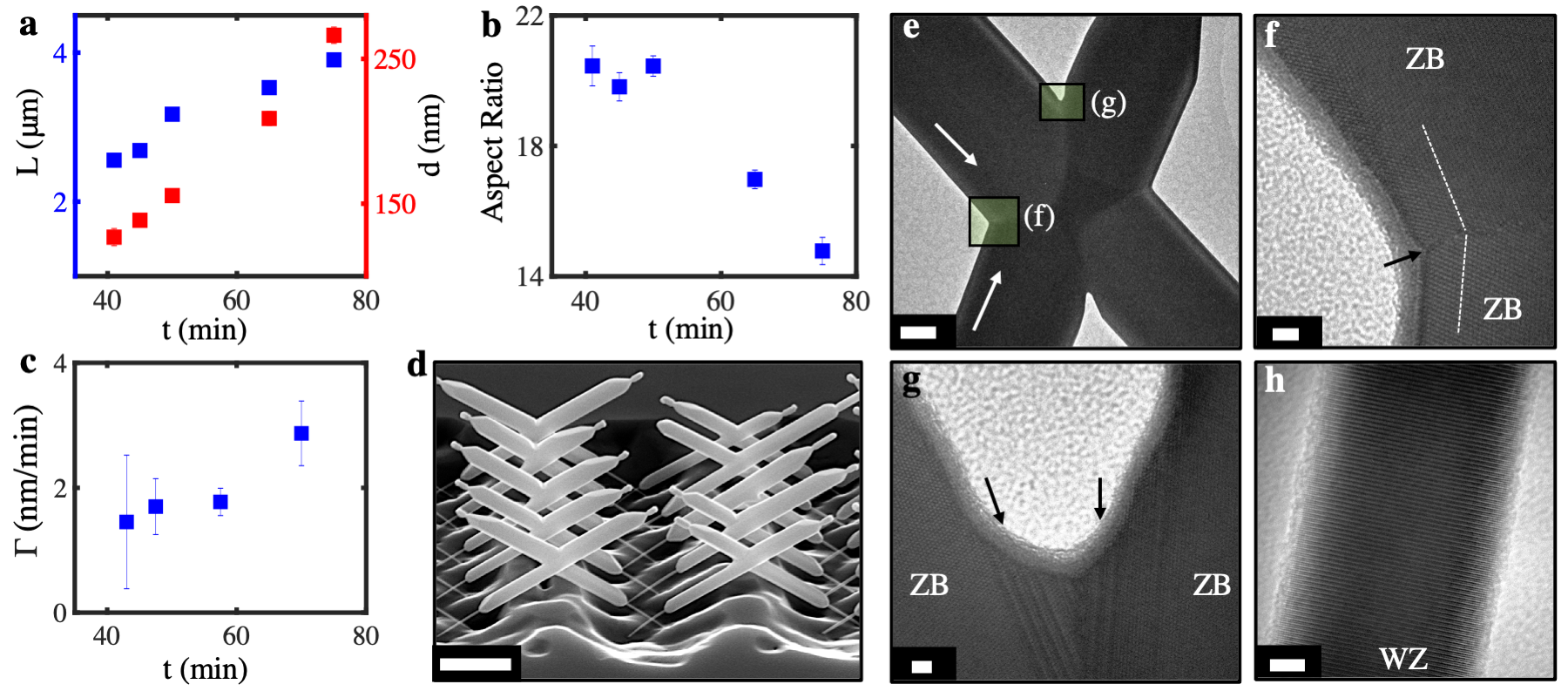}
\vspace{0.2cm}
\caption{\textbf{Growth dynamics and structural analysis of InSb nanocrosses.} \textbf{a}, Length and diameter of InSb nanowires at different growth times. Here, the time for growing the InAs stem is excluded. \textbf{b}, Aspect ratio as a function of InSb growth time. \textbf{c}, Radial growth rate data extracted from panel (a). \textbf{d}, Tilted SEM image of the high yield InSb nanocrosses, where NWs are merged with the radial growth. \textbf{e}, Representative TEM of a nanocross. Zinc Blende crystal structure is maintained before and after merging. Arrows indicate growth direction. \textbf{f-g}, High resolution TEM images of regions highlighted in panel (e). In the acute corner (g), few stacking faults are observed (indicated by arrow) in the merged section. \textbf{h}, TEM image of the pure wurtzite InAs stem. Scale bars in panels (d-h) are, (d): 1 $\mu \mathrm m$, (e): 100 nm, (f-g): 2 nm, (h): 5 nm.}
\label{fig2}
\end{figure*}


To find the optimal parameters for NC formation five growths of different growth-times between  41 and 75 minutes were performed, and Fig.\ \ref{fig2}\textbf{a} shows the measured average InSb lengths and final diameters from NWs grown in trenches (with $\mathrm{\Delta x = 4.5}$ $\mu$m and pitch size of 970 nm between Au catalysts). Both length and diameter increase with time but while the diameter more than doubles in this time interval, the length only increases by a factor of 1.5. This trend is also evident from the decaying aspect ratio extracted in Fig.\ \ref{fig2}\textbf{b}. The InSb NW growth is initiated by axial growth via the liquid-solid transition, with an initial diameter $d_{0}$, which is determined by the volume and contact angle of the Au particle at the point of supersaturation with In and Sb. After some time, the NW length becomes comparable to the incorporation limited diffusion length $\mathrm{\lambda}$ and adatoms cannot reach to the Au particle, initiating radial growth on the side facets. In general, radial growth can be divided into three stages: no radial growth ($\mathrm{L<<\lambda}$), transition stage ($\mathrm{L\approx\lambda}$), and constant radial growth ($\mathrm{L>>\lambda}$). The extracted radial growth rate is presented in Fig.\ \ref{fig2}\textbf{c} from which is can be seen that the main part of the InSb growth resides in the transition stage for the studied growth times from 43 to 70 minutes (see Fig.\ \ref{fig2}\textbf{c}). Note that, the growth parameters were chosen to promote radial growth, rather than long NW with high aspect ratio, as this is required for the formation of high quality NCs. 

Extrapolating the diameter in Fig.\ \ref{fig2}\textbf{a} to $t=0$, we estimate a diameter of the growth front of $\sim 100 \, \mathrm{nm}$ and choosing $\Delta \mathrm y \approx 185 \, \mathrm{nm}$, the growth fronts is expected to pass uninterrupted at $t_\mathrm{MP}$. The length of the combined InAs/InSb NW at the meeting point $L_\mathrm{MP} = \Delta{x}/2\cos{\theta}$ with $\mathrm{\theta= 35^{\circ}}$ being the inclination angle of the (111)B facet, is determined by the position of the Au catalyst particles and was $2.8 \, \mu \mathrm{m}$ for our substrates. The InAs stem was grown to a length of $\sim 1.5 \, \mu \mathrm m$ leaving $1.3 \, \mu \mathrm m$ for the InSb to reach the meeting point (MP). As seen in Fig.\ \ref{fig2}\textbf{a}, the NWs reach this point after a growth time shorter than $t_\mathrm{MP} \lesssim 30 \, \mathrm{min}$ having then a diameter $d(t_\mathrm{MP}) \sim 100 \, \mathrm{nm}$, and therefore, the NWs pass each other at the MP as separate structures. Figure \ref{fig2}\textbf{d} shows an example of a growth performed for $75 \, \mathrm{min}$ leading to high throughput NCs formation. Here, the final length of the InSb NWs were $\sim$ $\mathrm{4}$ $\mu$m with diameter of $\sim$ $\mathrm{265\pm 6}$ nm and the radial overgrowth results in clear joining of the NW pairs into four-terminal NCs.

Structural TEM characterization is presented in Fig.\ \ref{fig2}(\textbf{e-h}) and further details are provided in Supporting Information S1. Figure \ref{fig2}\textbf{e} shows the overall structure of a NC (grown along the direction of the arrows) and confirms that all the four arms along with the merged region maintain the same crystal structure throughout. The high contrast structure in the middle region is a consequence of the difference in thickness of the structure. No stacking-faults or crystal defects were observed in the NC arms. Figure \ref{fig2}\textbf{f} shows a high-resolution image of the obtuse corner between the NWs. The ZB crystal phase is maintained also in the structure closest to the surface, which is grown radially from the side facets and responsible for the merging of the crystals. A twin-plane is observed at the point of merging, which we attribute to misalignment or different stacking order of the two phases. Correspondingly, Fig.\ \ref{fig2}\textbf{g} shows a high resolution image of the acute corner where multiple stacking faults are observed in the surface layers growh radially (arrows) presumably also caused by misalignment of the two crystals. These, however, only appear in the radially grown part of the crystal and do not propagate into the core of the wires. Figure \ref{fig2}\textbf{h} shows the high resolution (HR)-TEM image of wurtzite (WZ) InAs stem without any disorders. InAs stems get thinner with longer InSb growth time due to the As decomposition. \\\\


\begin{figure*}[ht!]
\vspace{0.2cm}
\includegraphics[width=18cm]{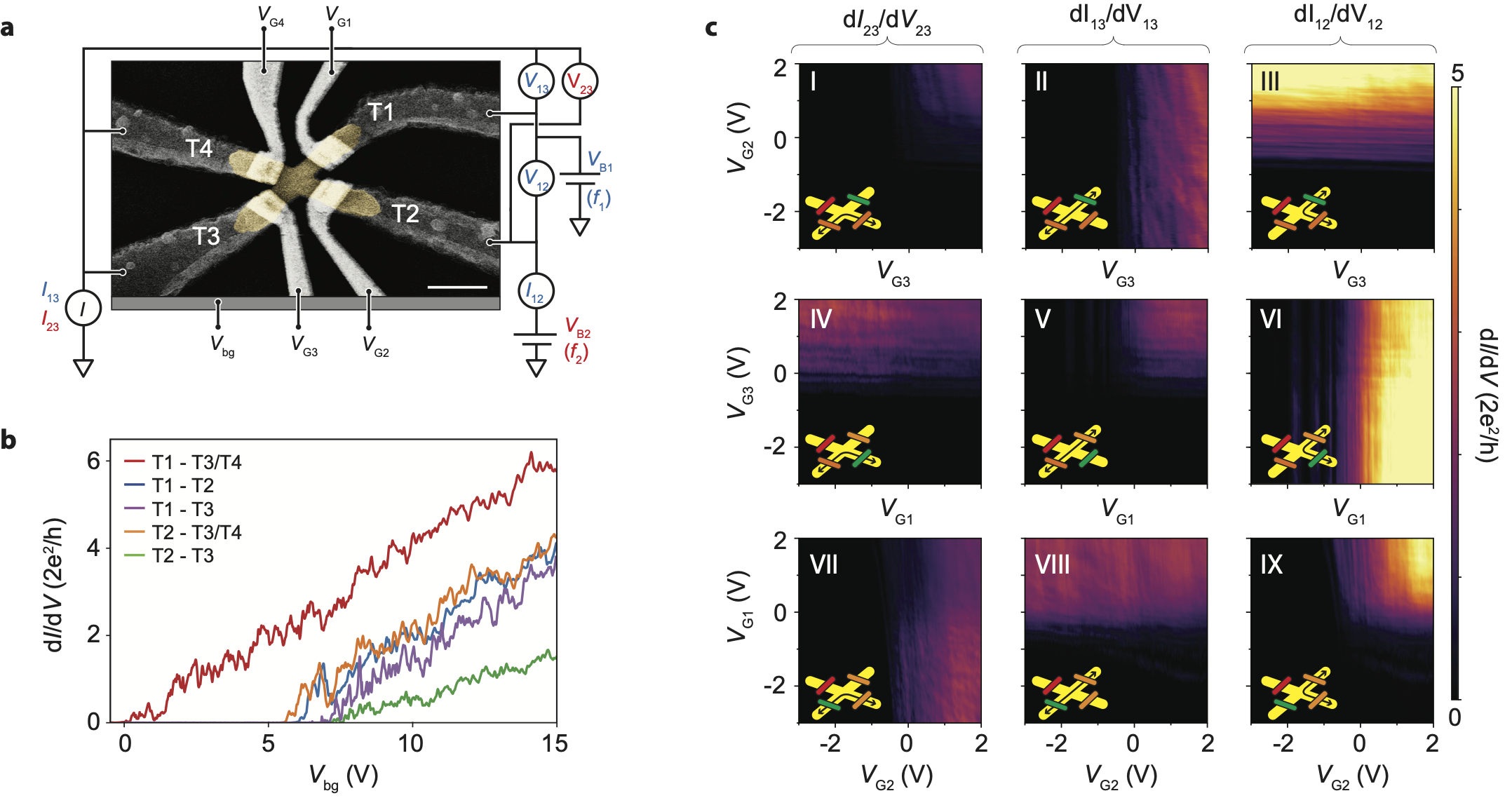}
\vspace{0.2cm}
\caption{\textbf{Individual gate control of a multi-terminal nanocross device} \textbf{a}, Scanning electron micrograph of a InSb nanocross device with four contacted terminals (T1-T4) and four individual top gates partly overlapping both contact metals and the InSb. The gates are biased at $V_\mathrm{G1},...,V_\mathrm{G4}$ and the measurement circuit used to simultaneously measure the conductances $G_\mathrm{12}, G_\mathrm{13}$, and $G_\mathrm{23}$ are shown. The two \textit{ac} biases $V_\mathrm{B1}$ and $V_\mathrm{B2}$ are applied with the two incommensurable frequencies $f_1$ and $f_2$ and lock-in detection of the currents $I_{12}$, $I_{13}$ and voltages $V_{12}$, $V_{13}$ are carried out at frequency $f_1$, while the rest is measured at $f_2$, as indicated by the colors. \textbf{b}, Two-terminal conductance vs. back gate potential of all possible connections through the cross. Top gates were grounded except for the green and purple traces where T4 was in pinch off ($V_{G4} = -4 \, \mathrm{V}$).  \textbf{c}, Response of the three conductances $G_\mathrm{12}, G_\mathrm{13},G_\mathrm{23}$ to tuning each pair of top gates for $V_{bg} = 10 \, \mathrm{V}$ and $V_{G4} = - 4 \mathrm{V}$. All 9 combinations follow the expectations for independent gate control and weak cross coupling.}
\label{fig3}
\end{figure*}

\textbf{Low Temperature Transport Characterization.} We now consider the low-temperature electrical properties of the MBE-grown InSb NCs. Previous studies of InAs \cite{Suyatin:2008,Heedt:2016,Rieger:2016,Kang:2013} and InSb NCs \cite{plissard2013formation,Fadaly:2017,Gazibegovic2017epitaxy} have confirmed transport both along and between the two merged NWs, where both ballistic transport at high fields \cite{Fadaly:2017} and phase-coherence \cite{Gazibegovic2017epitaxy} have been demonstrated. A main reason for using branched nanostructures is the potential to individually gate control nanowire branches, thus enabling e.g.\ measurements of the local density of states \cite{Zhang:2019}, a control of the effective size of the scattering matrix in coherent multi-terminal junctions \cite{riwar2016multi}, or eventually braiding Majorana zero modes in topological junctions \cite{Alicea:2011}. So far, however, only global electrostatic gating has been demonstrated. 

For device fabrication, NCs were located on the growth substrate using SEM and subsequently transferred using a manual micro-manipulator to highly doped Si substrates capped with $500 \,\mathrm{nm}$ of SiO$_2$. Ohmic contacts to each terminal (denoted $T1, T2, T3, T4$) were defined by e-beam lithography and deposition of Ti/Al metals. Subsequently, 10 nm of $\mathrm{HfO_{x}}$ was deposited using atomic layer deposition and four individual Ti/Au top gates (potentials $V_\mathrm{G1} ,V_\mathrm{G2},V_\mathrm{G3},V_\mathrm{G4}$) were defined on the NC terminals. The gates overlap the electrodes and $\sim 100$ nm of the exposed InSb, and their separation is $300-400 \mathrm{nm}$. Figure \ref{fig3}\textbf{a} shows an SEM micrograph of the device. The back-gate ($V_\mathrm{bg}$), acts globally and in particular at the center of the cross. Terminals T3 and T4 were connected on the chip and most measurements had $V_{G4} = -4 \, \mathrm V$, keeping terminal $T4$ in pinch-off.

Measurements were carried out either by a pair-wise two-terminal AC lock-in configuration with the unused terminals floating, or employing a three-terminal AC setup, as schematically shown in Fig.\ \ref{fig3}\textbf{a} allowing simultaneous measurements of all relevant combinations $G_{13}=dI_{13}/dV_{13}$, $G_{23}=dI_{23}/dV_{23}$, and  $G_{12}=dI_{12}/dV_{12}$ (See Methods). Measurements were performed in a dilution refrigerator at an electron temperature $\sim 20 \, \mathrm{mK}$. We note that Al is a superconductor at the measurement temperatures, however, no signatures of superconductivity were detected, presumably due to a disordered interface between InSb and the Ti/Al electrodes as further discussed below.

Figure \ref{fig3}\textbf{b} shows the back gate dependence of the conductance of individual connections T1 - T3/T4 (red), T2 - T3/T4 (orange), and T1 - T2 (blue) measured with the remaining terminals floating and top-gates grounded. For the purple and green traces the connections (T1-T3) and (T2-T3) were measured with $V_\mathrm{G4} = -4 \, \mathrm{V}$ to disconnects T4. In all cases, conductance increases with more positive $V_\mathrm{bg}$, as expected for a $n$-type semiconductor and the conductance values for $V_\mathrm{bg} = 15 \, \mathrm  V$ is relatively high for such devices indicating low resistance Ohmic contact. 
The $G(V_\mathrm{bg})$ traces exhibit pronounced reproducible oscillations, which are attributed to conductance fluctuations due to phase coherence as commonly observed in NW devices at low temperatures \cite{Doh:2005}.

\begin{figure*}[ht!]
\vspace{0.2cm}
\includegraphics[width=18cm]{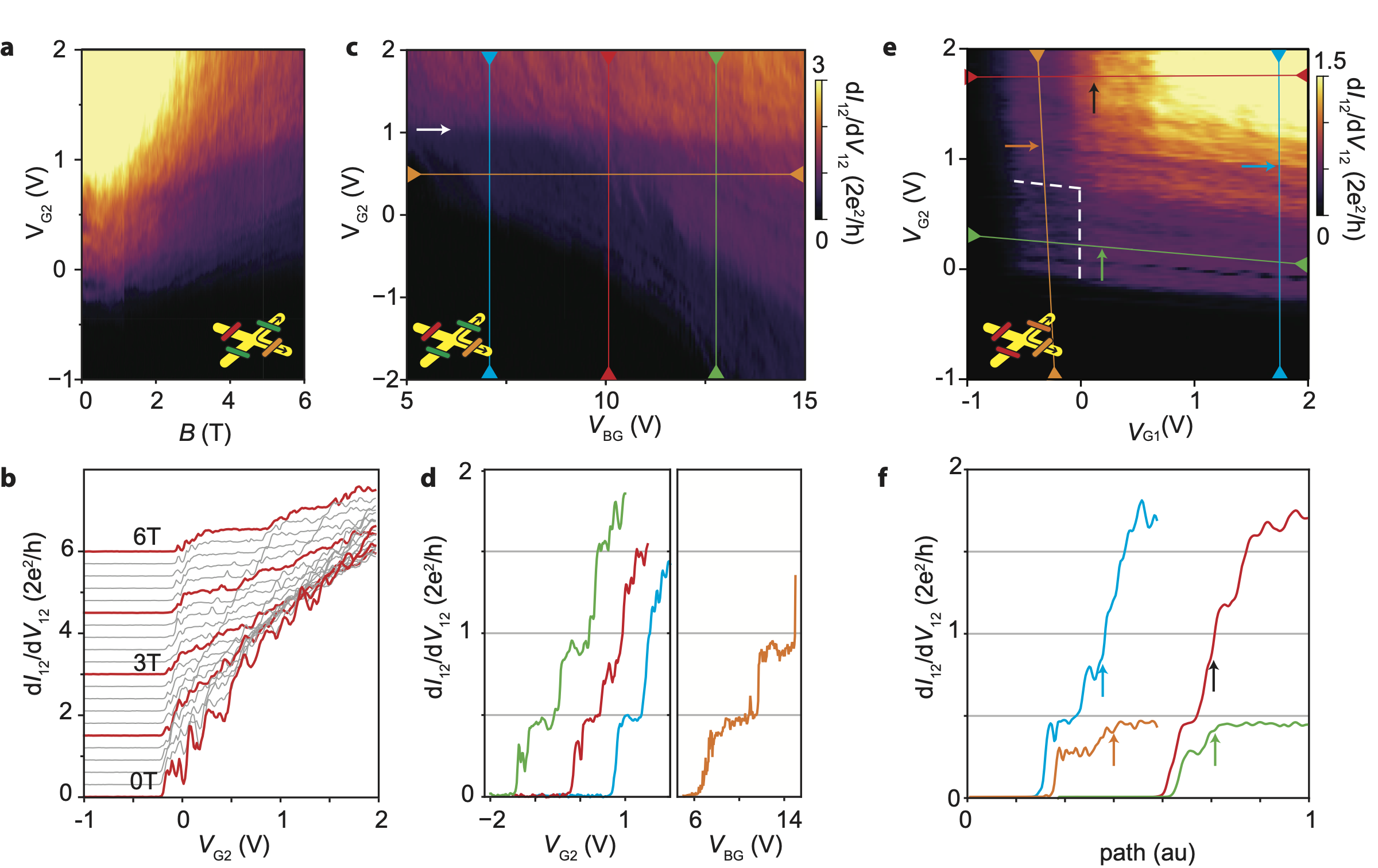}
\vspace{0.2cm}
\caption{\textbf{Ballistic transport through InSb nanocross.} \textbf{a}, $G_{12}$ vs.\ $V_\mathrm{G1}$ and $B$ for $V_\mathrm{bg} = 10 \, \mathrm V$. Corresponding line-cuts are shown in \textbf{b}, showing the development of quantized steps at $e^2/h$. \textbf{c}, As \textbf{a}, but measured vs.\ $V_\mathrm{bg}$ for $B=6 \, \mathrm T$. The positions of the line traces in \textbf{d} are shown with the corresponding colors with a offset of 0.5V. \textbf{e}, $G_{12}$ vs.\ $V_\mathrm{G1}$ and $V_\mathrm{G2}$ characterizing the serial connection through the two constrictions at $B=6 \, \mathrm T$ and $V_\mathrm{bg} = 10\, \mathrm V$. Line traces in \textbf{f} are indicated by colors, and arrows in \textbf{e} and \textbf{f} indicate corresponding positions along the line cuts. For clarity, red/green traces in \textbf{f} are horizontally off set with respect to blue/orange.}
\label{fig4}
\end{figure*}

\begin{figure*}[t]
\includegraphics[width =17 cm]{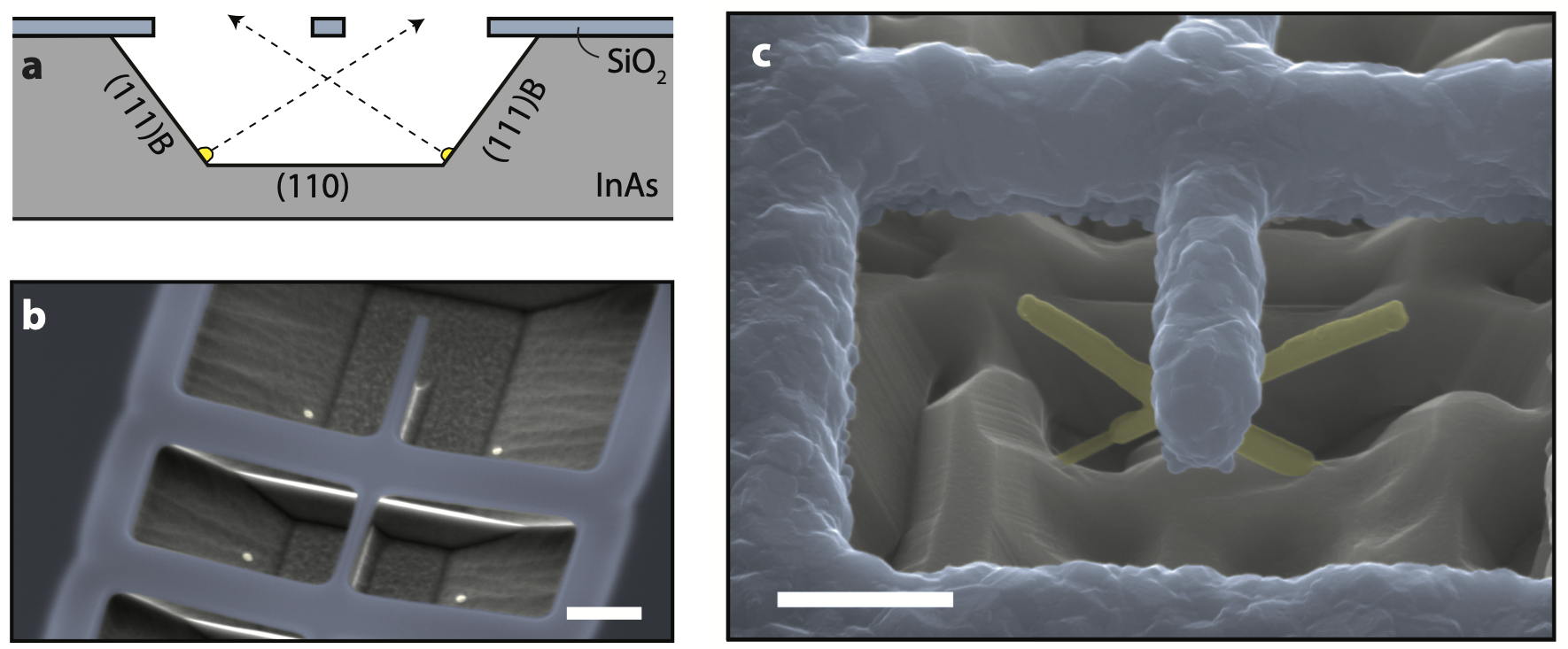}
\caption{\textbf{\textit{In situ} shadow epitaxy for hybrid semiconductor/superconductor nanocrosses} \textbf{a}, Schematic cross section of the trench substrate with a suspended bridge before growth. The growth direction of the two nanowires are indicated by dashed arrows. \textbf{b}, Tilted scanning electron micrograph of the substrate before growth. The $\mathrm{SiO_x}$ section is colored blue and Au catalyst particles beneath the bridge are seen as white dots. \textbf{c}, SEM image of the nanocross growth below the shadow bridge. Substantial overgrowth is observed both on the $\mathrm{SiO_x}$ mask and in the trench. During \textit{In situ} hybridization, Al deposition is aligned in a way that the bridge shadows the middle of the nanocross. Scale bars in (\textbf{b}) and (\textbf{c}) are 1$\mu \mathrm{m}$.}
\label{fig5}
\end{figure*}

The threshold for T1-T3/T4 is $V_\mathrm{th} \sim 0 \, \mathrm V$, while it is shifted to 6V for T2-T3/T4, showing that $V_\mathrm{bg}$ acts non-uniformly on the terminals activating T2 at higher $V_\mathrm{bg}$. Also, closing T4 ($V_\mathrm{G4} = -4 \, \mathrm V$, purple trace), shifts $V_\mathrm{th}$ to 6V suggesting that for $V_{bg} \lesssim 6 \, \mathrm V$ transport is dominated by a local path T1-T4. Close to pinch-off transport is thus not spatially uniform. Except for T2-T3, however, the transconductances are similar for all combinations $dG/dV_\mathrm{bg} \sim 30 \mu \mathrm{S/V}$, and estimating the back-gate capacitance by a simplified model of a NW above a planar back gate \cite{wunnicke2006gate}, we find a field effect mobility $\mu$  of $\mathrm{\sim}$ 700 $\mathrm{cm^2/(Vs)}$. This rough estimation disregards field focusing by the cross geometry and screening from electrodes and top-gates, but the value is comparable to the values reported for individual InSb NWs \cite{gul2015towards}.

Turning now to the gate control and capacitive cross-coupling of individual arms of the NC, we fix $V_\mathrm{bg} = 10\,\mathrm V$, and Fig.\ \ref{fig3}\textbf{c} shows the three simultaneously measured conductances $G_{23}, G_{13}, G_{12}$ (columns) vs.\ different pairs of top gates. The non-swept gates were kept at 2V, thus keeping the corresponding terminal open; the configuration for each panel is indicated by the icons. All top gates show a consistent threshold at $\sim 0 \, \mathrm V$ and the maps qualitatively follow the expectations of a transistor network with little cross-coupling between the gates. Consider for example panels I-III, where $V_{G2}$ (vertical axis) modulate the connection between T2 and T3 (panel I) and between T1-T2 (panel III), while $G_\mathrm{13}$ is unaffected (panel II). Also, for panels I, V, IX on the diagonal, which show $G_{nm}$ as a function of $V_\mathrm{Gn}$ and $V_\mathrm{Gm}$, finite conductance is only obtained when both gates are at positive potential. Cross-coupling due to the geometric proximity distorts conductance features from horizontal/vertical and the largest effect is observed between $V_\mathrm{G1}$ and $V_\mathrm{G2}$ in panel IX exhibiting a $\sim 12 \%$ cross-coupling. The main result of Fig.\ 3 is thus to establish the possibility for individually controlling the legs of the NC.

Although the gates act locally on the terminals of the cross and transport is coherent at low temperature, the measurements in Fig.\ \ref{fig3}\textbf{c} do not show clear quantized conductance presumably due to quantum interference and scattering dominating the transport. Previous studies of single nanowires \cite{Weperen:2013} and globally gated NCs \cite{Fadaly:2017} have found that a magnetic field can suppress scattering and enhance the signatures of ballistic transport. Figures \ref{fig4}\textbf{a,b} show $G_{12}$  vs.\ perpendicular magnetic field $B$ and $V_\mathrm{G2}$ with $V_\mathrm{G1} = V_\mathrm{G3} = 2 \, \mathrm V$ (T1 and T3 fully open). $V_\mathrm{G2}$ thus control the main barrier for transport and for $B \gtrsim 1 \, \mathrm T$, $G_2(V_\mathrm{G2})$ increases in discrete steps of $e^2/h$, the conductance of single, spin-split one-dimensional channels below the gate. This is further emphasized by the line-traces in Fig.\ \ref{fig4}\textbf{b}.

To study further the properties of the local barrier on the NC, Fig.\ \ref{fig4}\textbf{c} shows $G_{12}$ vs.\ $V_\mathrm{G2}$ and $V_\mathrm{bg}$ for $B= 6 \mathrm T$. See Supporting Information S7(\textbf{d,e}) for the remaining gate combinations. In all cases, regions of quantized conductance are clearly observed confirming the ballistic 1D nature of the gate-defined constrictions of the NC terminals. Two distinct transitions are distinguishable; one nearly independent of $V_{bg}$ (arrow in panel (c)), and others which are modulated by both $V_\mathrm{bg}$ and $V_{G2}$. We attribute these to spatially separated transport paths both acting as quantum point contacts; one located close to the top gate and thus unaffected by $V_\mathrm{bg}$ and the other closer to the bottom of the semiconductor terminal and thus affected by both gates. Figure \ref{fig4}\textbf{d} shows extracted traces at the positions in Fig.\ \ref{fig4}\textbf{c} indicated by arrows, highlighting the flatness of the plateaus and the effective doubling of the step height when simultaneously crossing transitions from both families (blue curve). Understanding of such complexity in the effective device geometry, observed even when back-scattering is suppressed by the magnetic field is important for interpreting and utilizing nanowire networks for complex quantum devices. Further insight of individual paths could potentially be gained through detailed analysis of the properties at finite bias or utilizing additional local gates, however, these are beyond the scope of this study. 
The observation of quantized conductance shows that transport is ballistic at a length scale of the effective constriction induced by the top gates. Compared to previous studies, the possibility of introducing ballistic point contacts at the individual terminals allows further insight into the transport specifically through the crossing region. For two ballistic constrictions in series (QPC1 and QPC2), a fully diffusive transport in the intermediate NC will result in an ohmic addition to the total conductance $G_{t} = G_\mathrm{QPC1}G_\mathrm{QPC2}/(G_\mathrm{QPC1}+G_\mathrm{QPC2})$ while, in the case of a fully ballistic intermediate transport, $G_{t} = \mathrm{min}(G_\mathrm{QPC1},G_\mathrm{QPC2})$. To this end Fig.\ \ref{fig4}\textbf{e} shows a measurement at $B=6\,\mathrm T$ and $V_\mathrm{bg}=10\,\mathrm V$ of $G_{12}$ vs. $V_\mathrm{G1}$ and $V_\mathrm{G2}$ while keeping T3 and T4 in pinch-off ($V_\mathrm{G3}=V_\mathrm{G4}=- 4 \, \mathrm V$). This adds the two ballistic constrictions at T1 and T2 in series. Consider first the red trace in panel \textbf{f}, extracted at $V_\mathrm{G2} \sim 1.75$. In this case, T2 is  fully open, and the total conductance increases step wise upon gradually opening T1. The green trace shows a similar situation except T2 is now maintained at the first plateau. Interestingly, upon opening up T1, the conductance first shows a plateau at $\sim 1/2 \times e^2/2$, as expected for an Ohmic (diffusive) addition of the two constrictions. However, upon further increasing $V_\mathrm{G1}$, the conductance increases to $e^2/2$ and -- as expected for ballistic transport between the two terminals -- stays constant while increasing the number of channels in T1. A similar behavior is observed in the reverse situation, where T1 is open and channels are gradually added to T2 (blue and orange traces in Fig.\ 4\textbf{f}). Thus, transport between the two constrictions, i.e.\ through the merged NWs, appears ballistic except close to pinch off. The corresponding results from the other combinations of terminals are presented in Supporting Information S8 consistent with this scenario. A plausible scenario is that the first modes are located closer to the surface and thus subject to stronger scattering than at higher densities, where transport occur through the bulk of the NC. \\

\textbf{\textit{In situ} Shadow Patterning of Nanocrosses}. The scattering in the junction, the need for high magnetic fields to observe quantized conductance of the constrictions, and the absence of proximity induced superconductivity from the evaporated Al leads, may be traced back to disorder related to the device processing. Replacing evaporation and post processing with growth of epitaxial superconductors is a well established method for avoiding disorder-related degradation of device performance \cite{Krogstrup:2015}. However the fragility of InSb is a challenge for devices based on InSb/Al epitaxial hybrids as all known  etchings of Al also severely damages the InSb semiconductor and degrade device performance. Recently, \textit{in situ} “shadow” approaches \cite{Carrad:2020,Gazibegovic2017epitaxy,krizek2017growth,khan2020highly,Heedt:2020} have been developed allowing patterning of epitaxial superconductor growth, yielding reproducible transport characteristics and observations of ballistic transport at $B=0\,\mathrm T$ for single nanowires. These \textit{in situ} shadow concepts, developed for single NWs, are however, incompatible with the NC geometry, where controlled shadows of the central junction region is required for most applications. Figure \ref{fig5} demonstrates a new adaptation of the approach in Ref.\ \cite{Carrad:2020} to accommodate the NC geometry. The substrates were prepared with $\mathrm{SiO_x}$ shadow structures suspended over the trenches. The design features four 1$\mu$m wide strips bridging each trench. These support a 150 nm wide cross-bar suspended along the middle of the trench. The cross-bar act as a shadow-structure for the junction of NC grown to have their merging point below the substrate surface plane. Figure \ref{fig5}\textbf{a} shows a schematic side view of the substrate and panel \textbf{b} shows a SEM micrograph of the substrate before the growth (see Methods for details). Figure 5\textbf{c} shows an example of a InSb NC grown below the cross-bar. Substantial overgrowth on the $\mathrm{SiO_x}$ mask is also observed in this case, effectively widening the shadow bridge, and further optimization of growth parameters and/or bridge design is needed. The results, however, demonstrate the feasibility of this approach, and we expect that the electrical performance of the shadow crosses will increase similarly to the results reported for conventional nanowire devices fabricated by \textit{in situ} shadow techniques \cite{khan2020highly, Carrad:2020}. 

\section{Conclusion}

In conclusion, we have presented a detailed study of MBE grown InSb NCs. From the time-dependence of axial and radial growth of InSb NWs, we determined the combination of geometric and growth parameters optimal for NC growth featuring a coherent crystal structures. InSb NC devices were fabricated with individual electrostatic control of all terminals, each showing clear quantized conductance at low temperature and high magnetic fields. Analyzing the combined action of two point contacts connected in series through the cross, showed that inter-wire transport in the NC is quasi-ballistic except close to pinch-off, where signatures of diffusive transport occurs. Finally, we developed and demonstrated a shadow-technique allowing epitaxial growth of hybrid semiconductor/superconductor NCs with \textit{in situ} shadow junctions aligned to the NC intersection point. The approach can be applied for any choice semiconductor and superconductor, but is demonstrated here for InSb/Al hybrid crosses. The technique is expected to dramatically reduce disorder-related scattering and thus be an important step towards clean quantum transport in complex multi-terminal nanowire devices.

\section{Methods}

\textbf{Substrate Fabrication and Nanocross Growth.} InAs (100) 2-inch wafers are used for substrate fabrication. (111)B trenches are created on the substrate using  $\mathrm{H_2SO_4}$ and $\mathrm{H_2O_2}$ solution based wet-etching process, as discussed in \cite{khan2020highly}. Later, electron beam lithography is used for defining the position of the Au seed particles on the (111)B trenches with an offset for forming NCs. Post-exposure development is done using standard 1:3 MIBK: IPA  solution. Electron beam evaporator with a rate of 1 \AA{}/s is used for depositing Au thin layer. Subsequently, lift-off is performed using acetone dipping. Next, cleaning is done with 2 min of sonication with acetone, rinsing with IPA and milli-q water. Finally, 2 min oxygen plasma treatment is conducted to avoid resist residues on the substrate.

InSb NCs are grown using Veeco GEN II MBE system. Initially, substrate is annealed at 590 $^{\circ}$C with arsenic over-pressure. Subsequently, InAs stem is grown typically for 12 min (prior to NC segment), where As/In flux ratio is maintained 9.78, resulting the length of the stems are $\sim$ 1.3- 1.6 $\mu$m. Depending on the substrate geometry the growth time (length) of InAs stem is varied. Next, the As flux is terminated and Sb flux is introduced in the system maintaining continuous In flux. Consequently, InSb segments are grown on top of the stems using Sb/In flux ratio $\sim$ 4.3. The growth time of InSb segment is varied depending on the space between the catalysts and radial growth radial growth required to overcome this space to create NC, as discussed in the growth section above. Throughout the NW growth the substrate temperature is maintained to $\sim$ 447 $^{\circ}$C (set temperature).

For hybridization of the NCs, Al deposition is performed \textit{in situ} in the MBE growth chamber. After the NC growth, the substrate is cooled down to $\sim -$ 36$^{\circ}$C. Usually, 8-10 hours is the waiting period before the temperature is reached and growth chamber is ready for Al deposition. Low temperature limits the adatom diffusion length and thermodynamically drives to form continuous Al thin film on InSb NW. Upon the temperature is reached, the deposition angle is adjusted with the Al source, which confirms the shadow region in the middle of the NC. Before unloading, 15 min of oxidization is performed to create $\mathrm{AlO_x}$ passivation layer and avoid the Al dewetting issue in the elevated temperature. 

\textbf{Structural Characterization.} The morphology of the NCs are characterized by SEM. Crystal structures and intersection of the NCs are characterized by TEM and HR-TEM of FEI Tecnai T20 G2 (200 kV of acceleration voltage, Thermionic LaB6/CeB6 e-beam source, point resolution of 0.24 nm, line resolution of 0.14 nm, and STEM resolution of 1.0 nm).

\textbf{Electrical Measurements.} The electrical setup is shown in Fig.\ \ref{fig3}(a). Terminals 1 and 2 are \textit{ac} voltage biased with two incommensurable frequencies (indicated by blue and red colors), and both resulting \textit{ac} currents are measured using lock-in detection at terminal 3 yielding $G_{13}=dI_{13}/dV_{13}$ and $G_{23}=dI_{23}/dV_{23}$, where $V_{13}$ and $V_{23}$ are the local voltage drops measured separately with lock-in detection at the respective frequencies. The bias at $T2$ is applied at the back of a current amplifier which measures the current at frequency $f_1$ thus yielding $G_{12}=dI_{12}/dV_{12}$. 

\textbf{Fabrication of Shadow Bridges.} Shadow bridges were fabricated along the lines of Ref.\ \cite{Carrad:2020}: 150 nm of SiO$_x$ were deposited by plasma enhanced chemical vapor deposition (PECVD) on InAs (100) substrates and patterned by EBL and HF etching into the pattern of bridges supporting the cross-bar for shadowing. Subsequently, the InAs trenches were etched using a separate EBL step and the same wet etchant as discussed before. Catalyst particles were defined by EBL on the facets as shown in Fig.\ 1. Careful alignment is required to ensure a NC junction forming directly below the shadow bar.

\section{Supporting Information}

The Supporting Information is available at: \url{https://sid.erda.dk/share_redirect/B9S92R2aTL}

\section{Author contribution}

S.A.K, J-H.K. performed the MBE growth, material analysis and optimization supervised by P.K.; L.S, T.M. performed electrical measurements, analyzed the results, and developed the shadow concept supervised by T.S.J.; S.A.K, L.S. and T.S.J. wrote the manuscript with input from all authors.

\section{Acknowledgement}

S.A.K, J-H.K. and P.K. was funded by European Union Horizon 2020 research and innovation program under the Marie Sk\l{}odowska-Curie Grant No. 722176 (INDEED), Microsoft Quantum and the European Research Council (ERC) under Grant No. 716655 (HEMs-DAM). T.S.J was supported by research grants from Villum Fonden (00013157), The Danish Council for Independent Research (7014-00132), and European Research Council (866158). Authors thank to C. B. S\o{}rensen for the maintenance of the MBE system.

\bibliography{ref}

\end{document}